\documentclass[conference]{IEEEtran}
\IEEEoverridecommandlockouts
\usepackage{cite}
\usepackage{amsmath,amssymb,amsfonts}
\usepackage{algorithmic}
\usepackage{graphicx}
\usepackage{textcomp}
\usepackage{xcolor}
\def\BibTeX{{\rm B\kern-.05em{\sc i\kern-.025em b}\kern-.08em
    T\kern-.1667em\lower.7ex\hbox{E}\kern-.125emX}}

\usepackage{booktabs} 

\usepackage{listings}
\usepackage{array} 
\usepackage{ragged2e}
\usepackage{enumitem}
\usepackage[disable]{todonotes}
\usepackage{url}
\usepackage{wrapfig}
\usepackage{hyperref}
\usepackage{multirow}
\usepackage{soul}
\usepackage{titlesec}
\usepackage{flushend}
\usepackage[justification=centering]{caption}

\usepackage{censor}
\usepackage[most]{tcolorbox}
\usetikzlibrary{calc}

\usepackage{listings}
\usepackage{color}

\definecolor{dkgreen}{rgb}{0,0.6,0}
\definecolor{gray}{rgb}{0.5,0.5,0.5}
\definecolor{mauve}{rgb}{0.58,0,0.82}

\newcolumntype{L}[1]{>{\raggedright\let\newline\\\arraybackslash\hspace{0pt}}m{#1}}
\newcolumntype{C}[1]{>{\centering\let\newline\\\arraybackslash\hspace{0pt}}m{#1}}
\newcolumntype{R}[1]{>{\raggedleft\let\newline\\\arraybackslash\hspace{0pt}}m{#1}}
\newcolumntype{P}[1]{>{\centering\arraybackslash}p{#1}}

\makeatletter
\newcommand\footnoteref[1]{\protected@xdef\@thefnmark{\ref{#1}}\@footnotemark}
\makeatother

\titlespacing*{\subsubsection}{0pt}{1.0ex plus 1ex minus .2ex}{1.5ex plus .2ex}

\colorlet{punct}{red!60!black}
\definecolor{background}{HTML}{EEEEEE}
\definecolor{delim}{RGB}{20,105,176}
\colorlet{numb}{magenta!60!black}

\lstdefinelanguage{json}{
    basicstyle=\normalfont\ttfamily,
    numbers=left,
    numberstyle=\scriptsize,
    stepnumber=1,
    numbersep=8pt,
    showstringspaces=false,
    breaklines=true,
    frame=lines,
    backgroundcolor=\color{background},
    literate=
     *{0}{{{\color{numb}0}}}{1}
      {1}{{{\color{numb}1}}}{1}
      {2}{{{\color{numb}2}}}{1}
      {3}{{{\color{numb}3}}}{1}
      {4}{{{\color{numb}4}}}{1}
      {5}{{{\color{numb}5}}}{1}
      {6}{{{\color{numb}6}}}{1}
      {7}{{{\color{numb}7}}}{1}
      {8}{{{\color{numb}8}}}{1}
      {9}{{{\color{numb}9}}}{1}
      {:}{{{\color{punct}{:}}}}{1}
      {,}{{{\color{punct}{,}}}}{1}
      {\{}{{{\color{delim}{\{}}}}{1}
      {\}}{{{\color{delim}{\}}}}}{1}
      {[}{{{\color{delim}{[}}}}{1}
      {]}{{{\color{delim}{]}}}}{1},
}

\begin{document}

\title{A case study experience from microservice system testing: towards a test benchmark for the microservice community }
\title{Benchmarks for End-to-End Microservices Testing}

\author{\IEEEauthorblockN{Sheldon Smith}
\IEEEauthorblockA{\textit{Computer Science} \\
\textit{Baylor University}\\
Waco, Texas, USA \\
sheldon\_smith2@alumni.baylor.edu}
\and
\IEEEauthorblockN{Ethan Robinson}
\IEEEauthorblockA{\textit{Computer Science} \\
\textit{Baylor University}\\
Waco, Texas, USA \\
Ethan\_Robinson2@alumni.baylor.edu}
\and
\IEEEauthorblockN{Timmy Frederiksen}
\IEEEauthorblockA{\textit{Computer Science} \\
\textit{Baylor University}\\
Waco, Texas, USA \\
timmyfrederiksen@gmail.com}
\and
\IEEEauthorblockN{Trae Stevens}
\IEEEauthorblockA{\textit{Computer Science} \\
\textit{Baylor University}\\
Waco, Texas, USA \\
trae\_stevens1@alumni.baylor.edu}
\and
\IEEEauthorblockN{Tomas Cerny}
\IEEEauthorblockA{\textit{Computer Science} \\
\textit{Baylor University}\\
Waco, Texas, USA \\
tomas\_cerny@baylor.edu}
\and
\IEEEauthorblockN{Miroslav Bures}
\IEEEauthorblockA{\textit{Computer Science} \\
\textit{Czech Technical University, FEE}\\
Prague, Czech Republic \\
buresm3@fel.cvut.cz}
\and
\IEEEauthorblockN{Davide Taibi}
\IEEEauthorblockA{\textit{M3S Cloud Group} \\
\textit{University of Oulu}\\
Oulu, Finland \\
davide.taibi@oulu.fi}
}

\maketitle

\begin{abstract}
Testing microservice systems involves a large amount of planning and problem-solving. The difficulty of testing microservice systems increases as the size and structure of such systems become more complex. To help the microservice community and simplify experiments with testing and traffic simulation, we created a test benchmark containing full functional testing coverage for two well-established open-source microservice systems. Through our benchmark design, we aimed to demonstrate ways to overcome certain challenges and find effective strategies when testing microservices. In addition, to demonstrate our benchmark use, we conducted a case study to identify the best approaches to take to validate a full coverage of tests using service-dependency graph discovery and business process discovery using tracing.
\end{abstract}


\section{Introduction} \label{sec:introduction}

Microservices are the mainstream approach to building cloud-native systems. Microservice Architecture (MSA) prescribes that a system comprises independently deployable services that interact
\cite{lewis2014microservices}. A microservice should have a single responsibility, meaning they only manage one specific part of the organization's needs. Microservices enable the decentralized evolution of system parts and their selective scalability.
Creating a service based on MSA helps eliminate the dependency on certain technologies, enabling users to access service-specific infrastructure \cite{cerny2018contextual}. However, microservices are developed and evolved by independent and decentralized teams, despite the result that users see as one holistic product.

Ensuring that the overall system functions properly, no matter it's internal dependencies and decomposition, is vital. With the ever-growing complexity of the system and especially when decentralized, it is important to aim for complete system coverage, which is often challenging to accomplish. But only then could we uncover whether all system functions are assessed to proper and improper inputs. 

Alongside functional testing, it is vital to assess that the system responds properly to various amounts of users and requests in concurrent load. Such load testing can accompany functional tests while monitoring system response times.

Considering functional and load testing, we identified that the microservices community is missing a test suite benchmark that could be used for further research advancements and evolution in this field. Because of this lack, we devised a case study to show our progress in generating such a benchmark.

The central goal of our contribution is to provide the scientific community with a comprehensive test suite  for two well-established microservice systems. Specifically, we are focusing this work on functional and load testing. These testing types are both an important part of testing the overall functionality of these systems. 

We introduce the test suite benchmarks through a case study. We also identify and discuss multiple challenges that arose while testing these systems and present some of the best practices and fixes to common problems. We developed a "best-effort" complete set of tests covering the endpoints of the multiple microservices within each system. The test suite can be used by researchers to validate their work on microservice testing on a common test~suite. 

The remaining sections are organized as follows. Section 2 provides background on the microservices and tools. Section 3 covers our case study followed by the introduced benchmark in Section 4. Section 5 concludes the paper.

\section{Background}\label{sec:background}

\todo[inline]{Timmy note 5/31 Do we need to explain this in such detail, or can we assume a certain level of knowledge. Feels to me that this whole preamble section could be condensed into a sentence per paragraph and stuck into the different subsections.}
\todo[inline]{TOMAS, we need drastic cuts. \\I suggest to condense intro + A + B to a single paragraph \\ system benchmarks are good meat }
\todo[inline]{TOMAS, we need to select each word carefully and can remove the things that are unnecessary text while we keep the message and good understanding for the readers;\\perhaps we can reduce "Facing Initial Pitfalls"  }

Functional testing ensures that various system features work in accordance with specifications or expectations of the system functionality. The test cases are typically based on the specifications of the software components and each test corresponds to a given software requirement or feature. This helps ensure that the parts of the overall software suite work independently. Focusing on the software requirements ensures that the output is consistent with the end user's expectations.

To simulate user interaction with microservice systems, we consider the system as a black box and use its user interfaces unaware of internal details. To aid with such testing, various frameworks bring the ability to design testing scripts. For instance, the Selenium framework facilitates web-system testing. It provides a method of automated web browsers triggering events as if users made these. 

Functional regression testing checks the system after modification ensuring all system features, flows, and functionalities are working as in the previous version, or by a specification or expectation of correct functionality. It helps with the quality assurance of the system and assists in avoiding unintended changes in the system caused by haste push for changes.


Besides functionality, analyzing how a system behaves during various load conditions is also important. 
Load testing is the process of simulating system demand that tests behavior under various conditions. The term 'load' in this case refers to the rate or number of users and requests accessing the given system \cite{jiang2008automatic}. Load testing helps ensure that user actions on the system are stable by evaluating how the system reacts to various amounts of load. 

There are three main ways that load testing can be executed. These load generation techniques include utilizing real users to generate the load, using load drivers, or deploying the load tests on special platforms \cite{jiang2015survey}. Assessing the behavior of a system over multiple amounts of load can highlight areas of improvement, such as bottlenecks or areas of failure within the system. These issues can often only be discovered using load testing because they only occur or are visible under a certain amount of load on the system. 

Similarly, there are frameworks that can aid with test automation. For instance, Gatling \cite{gatling} can run a test script and generate comprehensive load reports.


Another key metric in software testing is test coverage. It helps to measure the amount of testing done over a given system. The coverage of the tests shows which parts of the system are being executed throughout the tests, measured by a ratio of the number of particular elements of the system covered by tests to the total number of these elements in~the~system. 

\subsection{Microservice System Benchmarks}
\label{subsec:benchmark}

To demonstrate functional and load tests for microservices, two well-established and community-based systems were used that provide a wide range of functionality that could be tested throughout a case study. The Train-Ticket \cite{trainticket} is based on the Java platform, and the eShopOnContainers \cite{eshop} uses C\#. This system selection allows us to illustrate different scenarios in our case study, resulting in a shareable test benchmark for these systems.

\begin{figure}[h]
    \centering
    \hspace{-1.3em}
    \includegraphics[width=1.02\columnwidth]{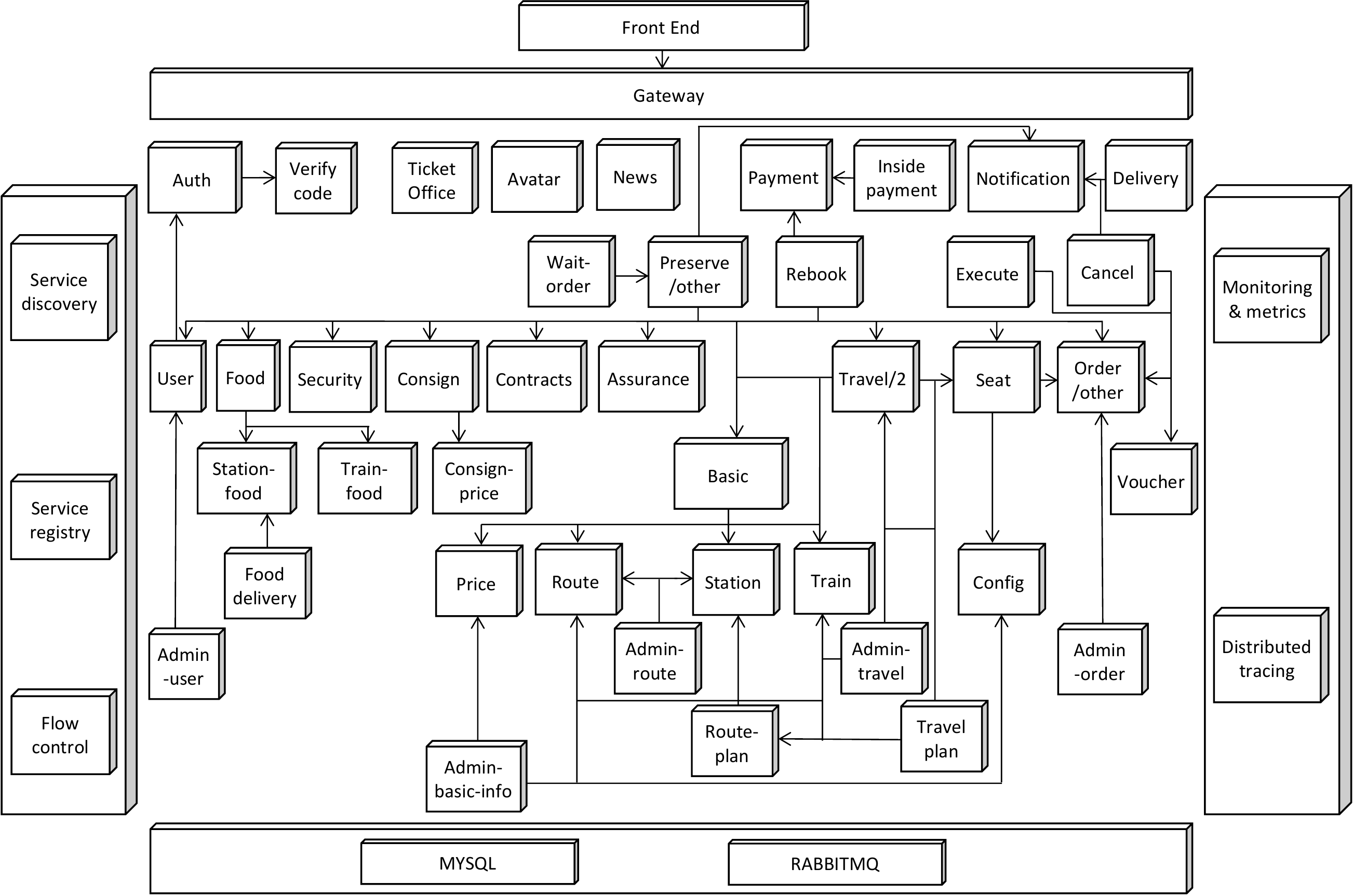}
       \caption{Train-Ticket Architecture \cite{trainticket}}
       \label{fig:train}
\end{figure}

{\bf The Train-Ticket benchmark} provides a train ticket booking system based on 47 microservices (as of version 1.0.0). Figure \ref{fig:train} shows the Train-Ticket architecture's general layout and structure. It shows how the front-end, monitoring system, and services interact within the system. This figure allows users to understand how each of the microservices within the Train-Ticket system depends on each other.

\todo[inline]{Sheldon (6/1): The comments on our paper mentioned that figures 1 and 2 were hard to read. How should we go about fixing that?}

The use cases involved with this microservice system can be broken down into user and admin actions. Certain actions are present for all users, such as the login system, while others depend on the type of user. 

There are six main \texttt{user} use cases within the Train-Ticket system. These cases are searching for a train, booking a ticket, updating one's consign, paying for the ticket, collecting the ticket, and entering the station. 

The \texttt{admin} use cases involve adding, updating, and deleting various elements of the ticketing system such as orders, routes, travel plans, users, contacts, stations, trains, prices, and configurations.

For our case study, we are using version 1.0.0 of Train-Ticket, which was released on August 9, 2022 \cite{trainticket}. This microservice system was created by the Fudan University CodeWisdom Team. The original goal for this system was to provide a benchmark system for railway ticketing \cite{zhou2018benchmarking}.

This system was created using a multitude of different programming languages and frameworks such as Java (Spring Boot, Spring Cloud), Node.js (Express), Python (Django), Go (Webgo), and MongoDB and MySQL for the databases \cite{trainticket}.


{\bf The eShopOnContainers microservice system} is a sample .Net Core reference application \cite{eshop}. The system is centered around providing various use cases involved in electronic shopping applications. The front-end for this microservice is split between two web applications: a traditional web app made using HTML and a Single Page Application (SPA) made through typescript and Angular 2. There is also a mobile app component to this microservice system.

\begin{figure}[h]
    \centering
    \includegraphics[width=\columnwidth]{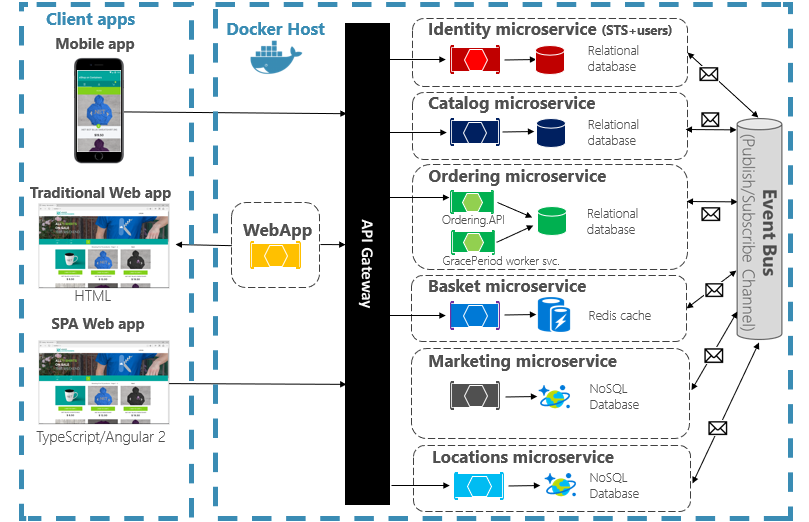}
       \caption{eShopOnContainers Architecture \cite{eshop}}
       \label{fig:eshop}
\end{figure}

The architecture of this application is cross-platform at the server and client-side. Figure \ref{fig:eshop} shows a layout of the interaction between the client apps and the Docker host. Within the Docker host there are multiple autonomous microservices with each service containing its own data or database. Different approaches to the structure of the microservices are used, such as CRUD and DDD/CQRS patterns. HTTP is the primary form of communication between these microservices and the client apps with which the user interacts \cite{eshop}. 

There are a multitude of use cases that the user can perform on this application. In order to interact the most with the functionality, the user must login with either the premade demo account or they can register a new account on the system. A user is able to filter the items based on multiple fields, which is not dependent on the user being login into the system. A \texttt{registered user} can add items to the cart, which can be updated on the cart page. Users can complete the checkout process to finalize the order or cancel the given order. Lastly, users can view the past orders that they've completed.

 For our case study we used version 5.0.0 of the eShopOnContainers microservice. This microservice system is provided as one of the reference apps by .NET Application Architecture \cite{eshop}; in addition, it has broad community contributions in its codebase repository.

\section{Case Study}\label{sec:use_case}

The goal of our  case study is to create an all-inclusive set of tests centered around the functionality of all the system endpoints, so as to  evaluate the system and its functionality under stress and  determine the behavior of well-established microservice systems.

Since we are focusing on functional and load testing for our case study, our test benchmark is centered around evaluating the response time of actions, the user interface interaction, and the overall functionality of the system. 

The functionality testing considers the system as a black box and is unaware of internal decomposition to microservices, which is often reflecting the user view. The (web-based) user interface testing benchmark helps ensure consistency throughout the execution of user actions.

The functionality testing through the user interface ensures that the system displays information correctly and consistently. If there is variability within the displayed information, it could greatly affect the user experience.

The load testing aims to identify possible places of bottleneck and slow down. This is important because user experience can be significantly effected by these issues. However, with load testing, we might consider the system to be a gray box and focus on middleware-exposed endpoints.


As mentioned previously, load testing microservice systems is vital to ensure that the various microservices react well to different amounts of load. This can help pinpoint areas of bottleneck and failure within the system.

Throughout the process of our case study, multiple challenges arose while testing these microservice systems. Within this case study we will present some of the best practices and fixes to common issues that could arise while testing. 

\subsection{Functional Regression Testing Case Study}

We wanted to ensure the correctness of the provided functionality of the microservice system by implementing an automated web-based test suite. We used Selenium for this purpose, and we were able to ensure the microservices function and display properly across various browser environments. We wanted to ensure that we implemented automated tests to thoroughly test both the microservices' functionality and design. For this case study, our goal was to provide a complete test suite that includes functional and user interface testing in order to closely simulate a user’s experience when utilizing the microservice systems. 

We used Selenium framework to write automated tests that simulated a user’s experience of the Train-Ticket and eShopOnContainers microservice systems.


\subsubsection{Study Design}
First, we manually created a comprehensive list of use cases within each microservice. Our approach to creating a list of comprehensive use cases was to record every possible action the user could take when interacting with the microservice and test the result of that action. We kept track of the use cases in a spreadsheet and marked each use case as the automated test was completed. Keeping a list of test cases helped us keep track of which use cases were fully tested and provided a good indicator of our progress until full coverage. This document has been shared along with our test benchmark on Zenodo \cite{zenodo_benchmark}.

\subsubsection{Study Procedure / Facing Initial Pitfalls}
We ruled off Katalon Web Recorder \cite{katalon} since the resulting scripts contained a lot of unmodularized, repetitive, and unreadable code. We proceeded manually with our own Selenium scripts. Runnable as JUnit tests for our automated web browser testing, we would test our work by using a Chrome web browser. This web browser allowed us to visualize how our code interacted with the microservices and allowed us to stop the execution and inspect the web browser while debugging. While using a Chrome web browser is good to debug and write the tests, using this method is not efficient when running multiple tests, since multiple web browsers are generated for each test and clutter the screen.

We grouped similar use cases together into tests, to reduce the number of tests run. Grouping the tests together helped reduce redundant steps and increased efficiency. After the creation of the tests, instead of running them on the Chrome web browser, we transitioned to running the tests on the HTML Unit WebDriver, which is a web driver without a GUI. It supports JavaScript and will simulate a web browser for testing with other frameworks, such as JUnit \cite{html_unit}. This also sped up the processing time for our JUnit tests. However, we wanted to make our tests even more efficient by parallelizing~them.

We utilized TestNG framework with the ability to parallelize the tests \cite{testng}. Moreover, TestNG allows the tester to have more control over the tests and is able to specify the number of threads the tests will execute on. We transitioned from using JUnit as our testing framework to using TestNG. After switching frameworks, we parallelized our tests to make them more efficient. After the parallelization of the Selenium tests, now the time it takes to run them was reduced dramatically.  

\subsubsection{Study Results}
\todo[inline]{Timmy note 5/31 I'm not a huge fan of all of these large tables of stuff, just because I don't think they necessarily convey anything crucial.}
\todo[inline]{Tomas I would keep them, they seem interesting to me if I want to use the benchmark}

Since our procedure was to create an all-inclusive set of tests of the user-provided functionality, we aimed to interact with every possible functionality exposed to the user in order to best simulate the user's experience. As a result of our functional regression case study, we ended up testing a total of 51 use cases for the TrainTicket system and a total of 26 use cases for the eShopOnContainers system, as shown in Table \ref{tab:selenium-test-count}.

 \begin{table}[h]
    \centering
    \begin{tabular}{|P{4cm}|P{4cm}|}
    \hline
        \textbf{Test Suite} & \textbf{\# of Selenium Use Cases}\\
    \hline \hline
        Admin TrainTicket & 33\\
        Client TrainTicket & 18\\
    \hline
        \textbf{Combined TrainTicket} & \textbf{51}\\
    \hline 
        \textbf{eShopOnContainers} & \textbf{26}\\
    
    \hline
    \end{tabular}
    \caption{Number of Selenium Tests per Suite\\Use Case Benchmark}
    \label{tab:selenium-test-count}
\end{table}

 Although we strove to fully implement every possible use case, there was one use case in the TrainTicket microservice system we could not fully test, which was the consign service. Although we fully wrote out the test for the consign service, we were unable to get that service deployed in the TrainTicket system. Therefore our test would work theoretically, but we were unable to verify its results.

The TrainTicket booking test suite encompasses all of the client-side use cases, due to their co-dependence on one another. The booking test suite involves using the regular search feature to find and book a ticket, pay for the ticket, change the ticket order, collect the ticket, and enter the station. The booking test suite also includes using the advanced search feature to find and book a second ticket, cancel that ticket, and finally delete all of the information added at the end of the test. All of these use cases are actions a client would typically take when booking and managing their ticket.

\begin{table}[h]
\centering
\begin{tabular}{|P{1cm}|P{2cm}|P{4.5cm}|}
\hline
\textbf{Test Suite}              & \textbf{Test}              & \textbf{Use Cases}\\
\hline \hline
\multirow{6}{*} & \multirow{2}{*}{Client Login} & Valid/Invalid Login,\\
                        &                          & Invalid User/Password, Logout\\
                        \cline{2-3}
                        & \multirow{4}{*}{Booking} & Book Economy, Book First Class,\\
                        &                          & Create/Save Contact,\\
 Client                 &                          & Book with Assurance, with Food,\\
                        &                          & with Consign, Cancel\\
                        \cline{2-3}
                        & \multirow{2}{*}{Order List} & Invalid Phone Number,\\ 
                        &                          & Update Consign, Pay for Ticket\\
                        \cline{2-3}
                        & Collect Ticket           & Collect Ticket\\
                        \cline{2-3}
                        & Enter Station            & Enter Station\\
                        \cline{1-3}
                        & \multirow{2}{*}{Admin Login} & Valid/Invalid Login,\\
                        &                          & Invalid User/Password, Logout\\
                        \cline{2-3}
                        & Order List               & Add/Update/Delete Order\\
                        \cline{2-3}
                        & Route List               & Add/Update/Delete Route\\
                        \cline{2-3}
                        & Travel List              & Add/Update/Delete Travel\\
                        \cline{2-3}
                        & User List                & Add/Update/Delete User\\
                        \cline{2-3}
 Admin                       & Contact List             & Add/Update/Delete Contact\\
                        \cline{2-3}
                        & Station List             & Add/Update/Delete Station\\
                        \cline{2-3}
                        & Train List               & Add/Update/Delete Train\\
                        \cline{2-3}
                        & Price List               & Add/Update/Delete Price\\
                        \cline{2-3}
                        & Config List              & Add/Update/Delete Config\\
\hline
\end{tabular}
\caption{Name of TrainTicket Selenium Tests\\Use Case Benchmark}
\label{tab:selenium-trainticket-test-name-new}
\vspace{-1em}
\end{table}

The eShopOnContainers checkout test suite involves populating the cart with an item, navigating to the checkout screen, and verifying the system needs the formatted information to proceed with the checkout and placing the order.

\begin{table}[!t]
\centering
\begin{tabular}{|P{2,5cm}|P{5,5cm}|}
\hline
\textbf{Test}              & \textbf{Use Cases}\\
\hline \hline
 \multirow{3}{*}        & Valid/Invalid Login,\\
    Login               & Invalid User/Password,\\
                        & Logout\\
 \cline{1-2}
 \hline
 \multirow{6}{*}{Registration} & Valid/Invalid Credentials,\\ 
                        & Missing Credentials,\\
                        & Unique Username,\\
                        & Matching Passwords,\\
                        & Invalid Expiration Date,\\
                        & Login\\
 \cline{1-2}
 \hline
 \multirow{2}{*}{Browse Pages} & Next Page,\\ 
                        & Previous Page\\
 \cline{1-2}
 \hline
 \multirow{3}{*}{Searching/Filtering} & Search By Brands,\\
                        & Search By Types,\\
                        & Multi-field Search\\
\cline{1-2}
 \hline
 \multirow{3}{*}{Update Cart} & Increment Cart (Save/No Save),\\
                        & Decrement Cart (Save/No Save),\\
                        & Remove Item\\
 \cline{1-2}
 \hline
 \multirow{2}{*}{Checkout} & Information Population,\\
                        & Valid/Invalid Checkout\\
\hline
\end{tabular}
\caption{Name of eShopOnContainers Selenium Tests\\Use Case Benchmark}
\label{tab:selenium-eshoponcontainers-test-name-new}
\vspace{-2em}
\end{table}


Table \ref{tab:selenium-trainticket-test-name-new} and Table \ref{tab:selenium-eshoponcontainers-test-name-new} lists all of the TrainTicket and eShopOnContainers Selenium tests. The number of tests for both TrainTicket and eShopOnContainers is far less than the amount of use cases covering each microservice, as depicted in Table I. Since we grouped together similar use cases, we ended up with far less tests than use cases.

The parallelization of the eShopOnContainers test suites resulted in faster execution of the tests. Before parallelizing our functional regression tests, we observed that the execution time would average about 25 seconds. After we implemented Html Unit and the TestNG testing framework, we observed that the execution time would average about 6-7 seconds, which was the execution time of the slowest test. 

Although our tests are still relatively small in size, our case study is evidence that parallelizing functional regression tests saves time, especially for bigger projects and testing suites.

\subsection{Load Testing}

Within our load testing case study, we wanted to focus on testing the response time of the multiple endpoints within the microservice systems. More specifically, we wanted to supply tests so that the community is able to see how these microservices response to a variety of different loads. We sought to provide a way for those interested to be able to see the various issues that could arise as more users access the system. This includes seeing the potential bottlenecks and response time of the system.

As elaborated in the background, we used Gatling (version 3.9.2) to create our load tests for both systems. 


\subsubsection{Study Design}

To track which endpoints have been tested within the microservice, we created a version control document to show the progress of our tests. Within the document, we listed out every endpoint we had found within the microservice system and additional information such as the specific microservice the endpoints were within, the testing file in which the endpoint was tested, and lastly, whether the endpoint was fully covered or not. This helped to provide a structure and plan for the tests moving forward. This allowed us to split up the testing files more efficiently, such as basing the tests on the microservice each endpoint was associated with or on a use case. This document has been shared along with our test benchmark on Zenodo \cite{zenodo_benchmark}.

To create this version control list of the endpoints, we needed to manually fill out all of the endpoints and services that they are a part of. This took a decent amount of time to create and keep track of which tests we had fully tested. It would have been more efficient and less reliant on human error to have an automated system that tracked endpoints that we tested by analyzing the test source code. This would have cut down a significant amount of time spent to make sure that the version control list and our tests were in sync. However, this process was out of the bounds of our case study, so we were unable to implement this system.

In addition to creating a list of endpoints, we also highlighted the main use cases within each microservice. This helped us get a better understanding of how users would normally interact with the microservices to help guide our tests to focus on the use cases that the user would perform. This list of endpoints is included on Zenodo \cite{zenodo_benchmark}.

\subsubsection{Study Procedure / Facing Initial Pitfalls}

We ruled out the recording tool provided by Gatling. Writing the tests manually allowed us to develop more concise and focused code that was outside the bounds of the recorder.

One issue that was prevalent early on in the process of load testing was ensuring proper authentication through the test. When a user logs into the given system, an authentication token is generated, which is used in the header of subsequent calls in order to authenticate that the user has valid access to that endpoint. However, within Gatling, this authentication token is often hard-coded into the header and thus becomes obsolete on ensuing requests. This causes an HTTP 403 forbidden response from the system, thus blocking the request. To solve this issue, we saved the authentication token that was in the response body after completing the login request. You can save a specific part of the response body by using the function \texttt{.check(jsonPath("\$.token")} \texttt{.saveAs("user\_token"))} in Gatling. This saves the token in a local variable called "\texttt{user\_token}." The variable can then be accessed using  \texttt{\$\{user\_token\}} in the header of the request to provide the valid authentication token.

Another important decision with load testing is how to deal with form parameters. Form parameters can be used in a variety of HTTP requests but are mainly associated with \texttt{post}, \texttt{put}, and \texttt{patch} requests since they normally require information to be sent to the endpoint. Throughout our case study for load testing, we used two main types of providing form parameters in Gatling: using the \texttt{formparam} function or the \texttt{.body(RawFileBody())} function. 

The \texttt{formparam} function is used for each parameter needed for the form. The format is \texttt{.formparam(parameterName, parameterValue)}. We found that this option is best when there is a low number of parameters. Using this function helps the tester easily see all of the parameters within the given endpoint call. Below is a sample post request showing the usage of the \texttt{formparam} function in Gatling. This example uses references to variables defined elsewhere in the test file.

\begin{lstlisting}
// Example Usage of RawFileBody() Function
.exec(http("Example Post Request for Login")
    .post("/api/v1/login")
    .formParam("Email", "${email}")
    .formParam("Password", "${password}")
    .headers(...))
\end{lstlisting}

The other option is using the RawFileBody() function, which can be used with a variety of file formats. We used this function by employing JSON files. We found this option is best suited for forms that require a large number of parameters. This is because you are able to contain all the parameters in one external file instead of having them spread over multiple lines within the test scenario. This option is also generally more dynamic than using the formparam function because you are able to use a variety of different file formats depending on which format suits the given request the best. This option involves supplying a JSON file within the structure of the project that maps the parameter names to each parameter's value. Below is a sample post request showing the usage of the \texttt{RawFileBody()} function in Gatling.

\begin{lstlisting}
// Example Usage of formparam Function
.exec(http("Example Post Request for Login")
    .post("/api/v1/login")
    .body(RawFileBody("login_form.json"))
    .headers(...))
\end{lstlisting}

\subsubsection{Study Results}

Before discussing the results from our load tests, it is important to note that load tests will give different results for different server capacities. In our case study, we deployed the microservice system through a Ubuntu server. We ran the tests on a Windows machine with a disk capacity of 1 TB, 32 GB of memory, and a 3.20GHz, 6-core CPU. For the deployed microservice system, we had a single insurance per each microservice in the system.

From our analysis of the Train-Ticket microservice, we outlined 240 endpoints across microservices. However, only 41 of those services contain endpoints. Throughout our case study, we were able to complete full coverage of those endpoints to the best of our understanding. We split up the tests into 26 Scala testing files. 

Tables \ref{tab:train_booking} and \ref{tab:eshop-checkout} give a breakdown of the response time of the respective system for a variety of users on the system for a certain use case. The columns show the number of users during the test, the number of requests that took more than 800ms, the total requests throughout the test, and the percentage of the requests that took over 800ms. This information is used to determine the stability of the use case given multiple levels of example load. Each load was applied to the system over a 30-second interval to isolate the load as the independent factor. In section \ref{sec:proposed}, we will discuss how we formulated these benchmark metrics.

Table \ref{tab:train_booking} shows the results of our load testing of the ticket booking use case in Train-Ticket. In this given use case, the user logs into the system, finds a train, and then books a ticket for that train. For the purposes of isolating the ticket booking use case, we silenced the login scenario requests so they would not show up in our results. 

Analyzing the results of the test covering 100 users, the system reacts well to this low load by having less than 1\% of the response be over 800ms. Loads of 500 and 1000 cause 7.3\% and 10.2\% of the system, respectively to fall above the 800ms threshold. However, these results still pass our measurement specifications, mentioned in Section \ref{sec:proposed}, because the system was able to handle the particular load since the percentage of requests over 800ms was below 20\%. The last tests covering loads of 2500 and 5000 users fall above these specifications thus showing that the system doesn't respond well to loads at and above these amounts.

\begin{table}[h]
    \centering
    \begin{tabular}{|P{1.5cm}|P{2cm}|P{2cm}|P{1.5cm}|}
    \hline
        \textbf{Number of Users} & \textbf{\# Greater Than 800ms} & \textbf{Total Requests} & \textbf{Percentage over 800ms}\\
    \hline\hline
        100 & 3 & 706 & 0.4\%\\
        500 & 259 & 3,530 & 7.3\%\\
        1000 & 721 & 7,060 & 10.2\%\\
        2500 & 3,813 & 17,650 & 21.6\%\\
        5000 & 15,249 & 35,300 & 43.2\%\\
    \hline
    \end{tabular}
    \caption{Train-Ticket Booking Ticket\\ Use Case Benchmark}
    \label{tab:train_booking}
    \vspace{-1em}
\end{table}

For our load testing of the eShopOnContainers microservice system, we split up the tests into 6 main Scala testing files. The use cases we lower for this system because the system mainly had use cases for one user type. We broke up the tests based on the use case that they covered.

Table \ref{tab:eshop-checkout} portrays the same information as Table \ref{tab:train_booking} but covers the results from the \textit{order checkout} use case in the eShopOnContainers microservice system. Within this use case, the virtual users log into the system, go to their cart, and check out. Similar to before we removed the requests from the login scenario from our results to isolate the checkout scenario.

The first few tests with 100 and 500 users respectively show a small amount of slow response time. This shows that the system remains stable during these low levels of load. This test benchmark shows that there is a dramatic increase in response time between 1,000 users and 2,500 users. A load of 1,000 users falls close to our prescribed measurement specification of 20\%. However, the system responds poorly in handling 2,500 and 5,000 users. Since a majority of requests take over 800ms to finalize, the system is unsuccessful in responding effectively to loads of over 1,000 users.
 
\begin{table}[h]
    \centering
    \begin{tabular}{|P{1.5cm}|P{2cm}|P{2cm}|P{1.5cm}|}
    \hline
        \textbf{Number of Users} &\textbf{ \# Greater Than 800ms} & \textbf{Total Requests} & \textbf{Percentage over 800ms}\\
    \hline\hline
        100 & 0 & 2,244 & 0.0\%\\
        500 & 52 & 11,253 & 0.5\%\\
        1000 & 4,276 & 22,479 & 19.0\%\\
        2500 & 35,952 & 56,343 & 63.8\%\\
        5000 & 75,910 & 112,593 & 67.4\%\\
    \hline
    \end{tabular}
    \caption{eShopOnContainers Checkout\\ Use Case Benchmark}
    \label{tab:eshop-checkout}
    \vspace{-1em}
\end{table}

Throughout both systems, we noticed that the microservice controlling the login scenario reacted specifically poorly to large amounts of load. This is likely due to the level of validation that goes into checking the credentials of the user. This highlights a key bottleneck within the system for users.

\section{Proposed Benchmark}
\label{sec:proposed}

\todo[inline]{Sheldon (6/1): Should we shorten some of the intro stuff to the benchmark and mainly focus on our definitions of the three components of our benchmark?}

With our case study, we provide the community with a test benchmark of the microservice systems that we evaluated. Creating a test benchmark allows for an easy way to show the community how systems respond to a variety of different tests. In addition, they can use the benchmark to simulate system used to perform dynamic system analysis, security assessments, resilience, or scale testing.

One key to test benchmarks is that they are repeatable. It can serve as a point of reference that other products and services can be compared against. These benchmarks can also be used to compare the past, present, and future software releases with their respective benchmarks. A clear evolution of the software results can be traced using these benchmarks.




Overall, there are three main benchmark components \cite{benchmark_info}. 
\begin{itemize}
    \item[1. ] \textit{Workload Specifications}: This area of a benchmark covers determining the type and frequency of requests to be submitted to the system under a given test. Within load testing this selecting the overall number of users over a given amount of time that the test will execute. The workload specifications for web browser testing involve evaluating the parallelization of tests.
    
    \item[2. ] \textit{Metric Specifications}: This component centers around determining which specific element of a given test will be used for evaluation. These metrics can be simply whether to overall test passed or fail, the response time of the requests, the efficiency of the tests, or some other metric.
    
    \item[3. ] \textit{Measurement Specifications}: The last main component of a testing benchmark is determining how to measure the specified metrics to evaluate the results. This determination is also denoted as the Service Level Agreement (SLA) criteria. This can involve a certain threshold for the number of requests passed a certain response time or whether all the tests passed or not.
    
\end{itemize}

Within load testing, it is important to keep the load as realistic to the normal workload of the system as possible. This will give a better understanding of how the system reacts under typical circumstances. Unfortunately, there is little data about the normal amount of users that access these microservice system. With this in mind, we decided to test a variety of different loads starting with a low base test of 100 users and going up to 5,000 users as our max. This helped define our workload specification to be 100 users, 500 users, 1,000 users, 2,500 users, and 5,000 users. 

To keep test execution consistent, these loads are executed over a 30-second period of time. The metric specifications that we use to measure the results were the response time of each request. With each test, we are able to divide the number of responses that took less than 800ms, between 800ms and 1200ms, and greater than 1200ms. 

It has been found that the most preferred response time for systems is 0.1 seconds or 100ms. However, the maximum limit of acceptable response time is normally set at 1 second or 1,000ms \cite{hamilton2023response}. With this in mind, we decided to set our acceptable response time at 0.8 seconds or 800ms since it lent itself well with the testing tool that we used. This lead us to determine that the specification of measurements for the load tests would be the percentage of requests that took over 800ms to complete. We chose to use a percentage instead of a specific number to help this metric scale better to tests with different loads. We decided that a given use case was able to handle the particular load if the percentage of requests that took over 800ms is less than 20\%


While writing regression tests for a web browser, it is important to try and make the tests run as efficiently as possible to reduce execution time. Since the tests will be run many times after changes or updates to the code, they need to have a low execution run-time. To cut down on the execution time, we used the TestNG framework to parallelize  the Selenium tests for the microservices. We decided to dedicate each test to its own thread to reduce the testing time as much as possible.

We grouped our regression tests by service, so each test would include many assertions about the various parts of the given system. We used the individual assertions as our metric specifications. The individual assertions and actions taken during the course of the test are the determining factor for evaluation.

For our regression tests, we measure the assertions by determining whether a collective test passed or failed. If one of the assertions within a test failed, it will cause the entire test to fail. We measure the success of the test only by if the entire test passed or failed. 

After completion, if our case study results in test benchmark for particular microservice systems, we share our set of tests to Zenodo \cite{zenodo_benchmark}. 
We share the test benchmark to allow other researchers and test industry engineers in the community interested in tracking the testing results to expand them.

\section{Conclusion}
\label{sec:conclusion}

This paper presents a novel test suit that can be effectively used as a benchmark for research on software testing in microservice-based systems. Our approach considered both functional regression testing and load testing. We selected two well-established microservice systems to create a test benchmark to provide the community. These benchmark systems were Train-Ticket and eShopOnContainers.

Our twofold approach to evaluating the performance of microservice systems helps highlight key areas of failure within these systems. The load testing aspect of our case study highlights areas within the endpoints that could cause a bottleneck or failure given different amounts of load. Understanding how the system responds to various amounts of load can help designers understand areas to help improve user experience. Our second testing aspect involves analyzing the system using functional regression testing. Specifically within this approach, we used automated web-system testing to ensure that the system displays correctly and consistently across various browser environments.

The contribution of this work to the community is (1) to provide an open-source example of automated functional regression tests and load tests for microservice systems, as previously published examples are not sufficient, and (2) to produce an initial set of comprehensive tests for a proposed benchmark of well-established microservice systems.

    
    
    



\bibliographystyle{IEEEtran}
\bibliography{bibliography}

\end{document}